\documentstyle[prb,aps,preprint,epsf]{revtex}
\begin{document}
\title{Electron Paramagnetic Resonance Linewidths and Lineshapes
for the Molecular Magnets Fe$_{\mathbf 8}$ and Mn$_{\mathbf 12}$}
\draft
\author{Kyungwha ${\mathrm Park^1}$ 
\and M.~A.\ ${\mathrm Novotny^2}$ 
\and N.~S.\ ${\mathrm Dalal^3}$ \and S.\ ${\mathrm Hill^{4}}$, 
and P.~A.\ ${\mathrm Rikvold^{1,5}}$}
\address{${\mathrm ^{1}}$School of Computational Science and 
Information Technology, Florida State University, Tallahassee, Florida 32306 \\
${\mathrm ^{2}}$Department of Physics and Astronomy, Mississippi 
State University, Mississippi State, Mississippi 39762 \\
${\mathrm ^{3}}$Department of Chemistry, Florida State University,
Tallahassee, Florida 32306 \\
${\mathrm ^{4}}$Department of Physics, 
University of Florida, Gainesville, Florida 32611 \\
${\mathrm ^{5}}$Center for Materials Research and 
Technology and Department of Physics, 
Florida State University, Tallahassee, Florida 32306}
\date{\today}
\maketitle
\begin{abstract}
We study theoretically Electron Paramagentic Resonance (EPR) 
linewidths for single crystals of
the molecular magnets Fe$_8$ and Mn$_{12}$ as functions of energy
eigenstates $M_s$, frequency, and temperature when a magnetic field 
along the easy axis is swept at fixed excitation frequency.
This work was motivated by recent EPR experiments.
To calculate the linewidths, we use density-matrix equations, 
including dipolar interactions and distributions of the uniaxial 
anisotropy parameter $D$ and the Land\'{e} $g$ factor. Our 
calculated linewidths agree well with the experimental data. 
We also examine the lineshapes of the EPR spectra due to local
rotations of the magnetic anisotropy axes caused by defects in
samples. Our preliminary results predict that this effect leads to 
asymmetry in the EPR spectra.
\end{abstract}



\begin{figure}
\begin{center}
\epsfxsize=4.0cm
\epsfysize=4.0cm
\epsfbox{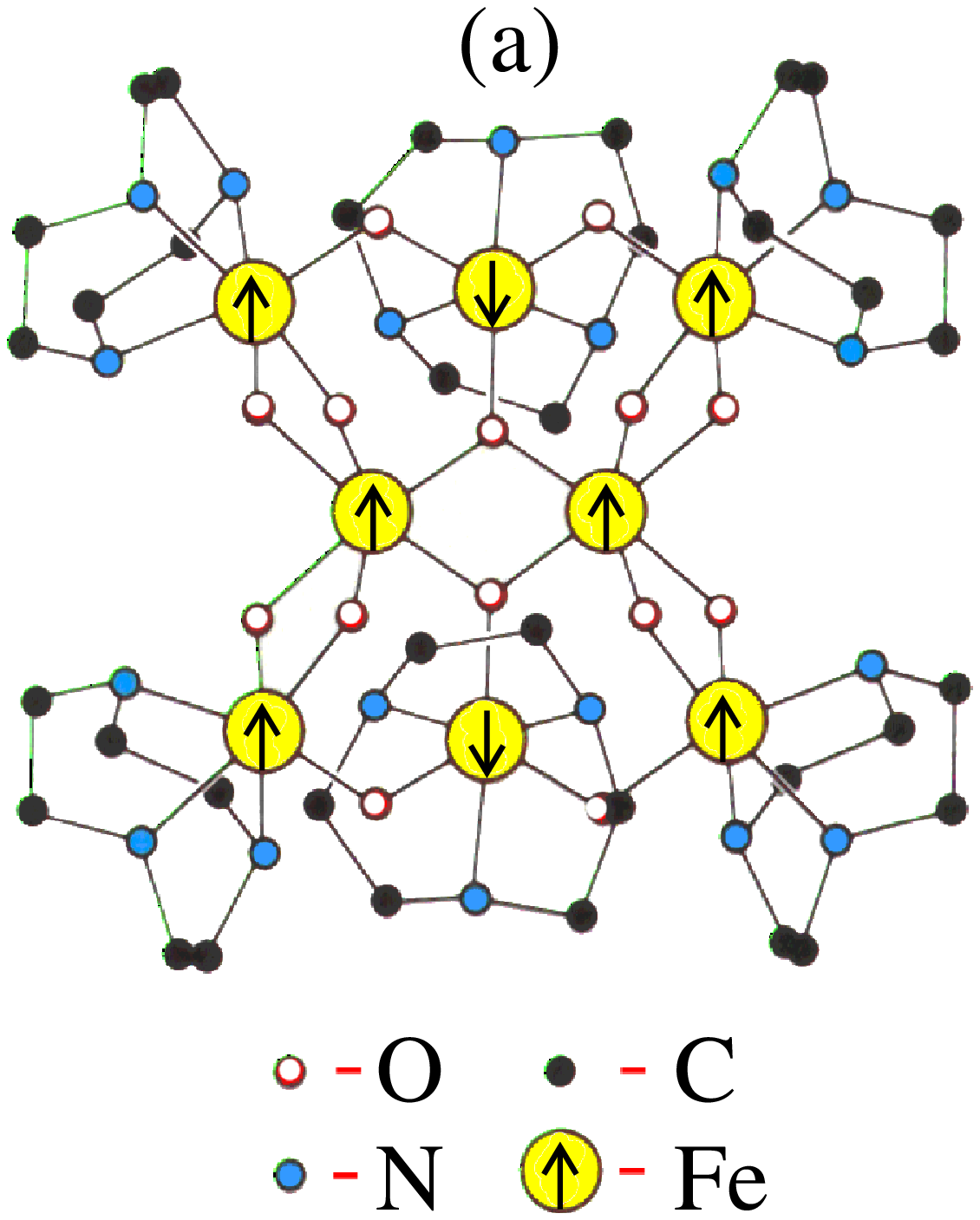}
\epsfxsize=4.0cm
\epsfysize=3.8cm
\epsfbox{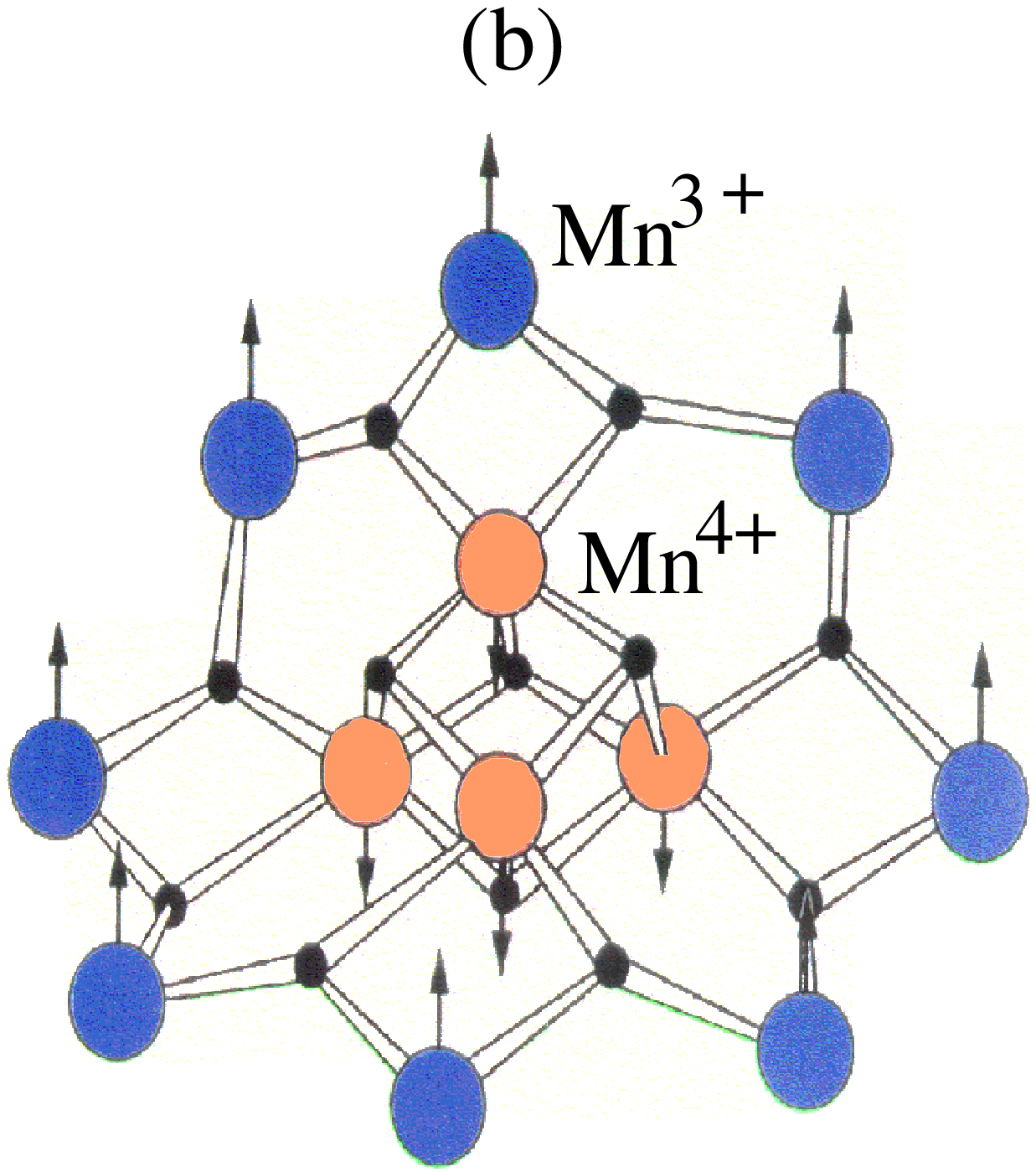}
\caption{Schematic diagram of the magnetic core of (a) the Fe$_8$ cluster
and (b) the Mn$_{12}$ cluster. Each Fe$_8$ cluster has D$_2$ symmetry, 
while each Mn$_{12}$ cluster has tetragonal symmetry. 
Both clusters have ground-state spin $S=10$
(for Fe$_8$, $S=(6-2)\times 5/2$, and for Mn$_{12}$, 
$S=8\times 2 - 4 \times 3/2$).}
\label{conf}
\end{center}
\end{figure}

Molecular magnets such as ${\mathrm Mn_{12}}$
\cite{LIS80} and ${\mathrm Fe_8}$ \cite{WIEG84} have recently 
drawn much attention because of the macroscopic quantum tunneling 
of their magnetizations 
at low temperatures \cite{CHUD98} and their possible applications
to quantum computing.\cite{LEUE00-3} 
These materials consist of many identical clusters (see Fig. \ref{conf})
with the same magnetic properties and characteristic energies.  
Each cluster has many different species of ions and atoms, with a
total spin angular momentum in the ground state of $S$$=$$10$.  
The clusters have strong crystal-field anisotropy and, thus,
a well-defined easy axis, 
and the magnetic interaction between different clusters is weak.

\begin{figure}
\begin{center}
\epsfxsize=5.5cm
\epsfysize=3.5cm
\epsfbox{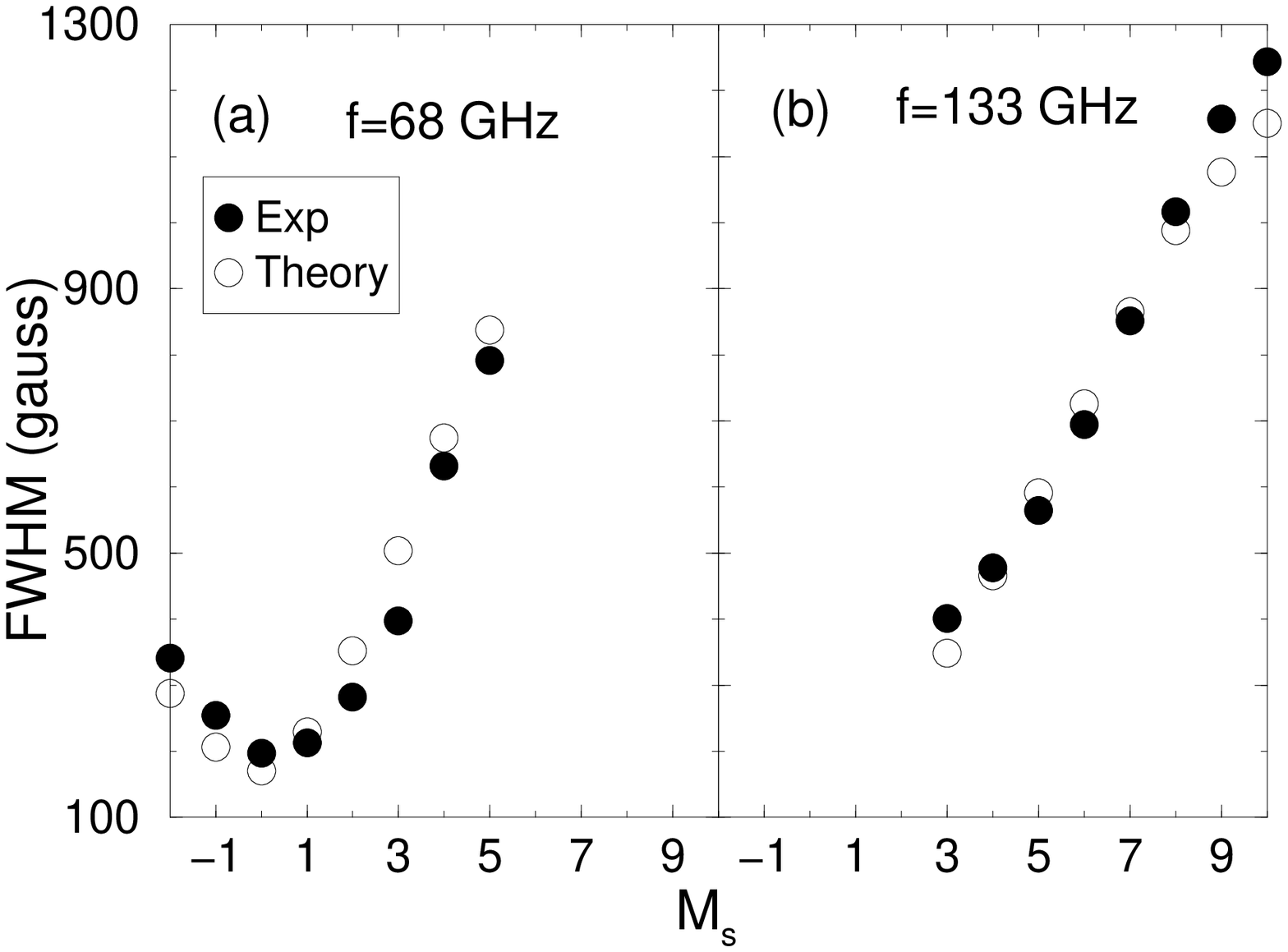}
\vspace{0.3cm}
\epsfxsize=5.5cm
\epsfysize=3.5cm
\epsfbox{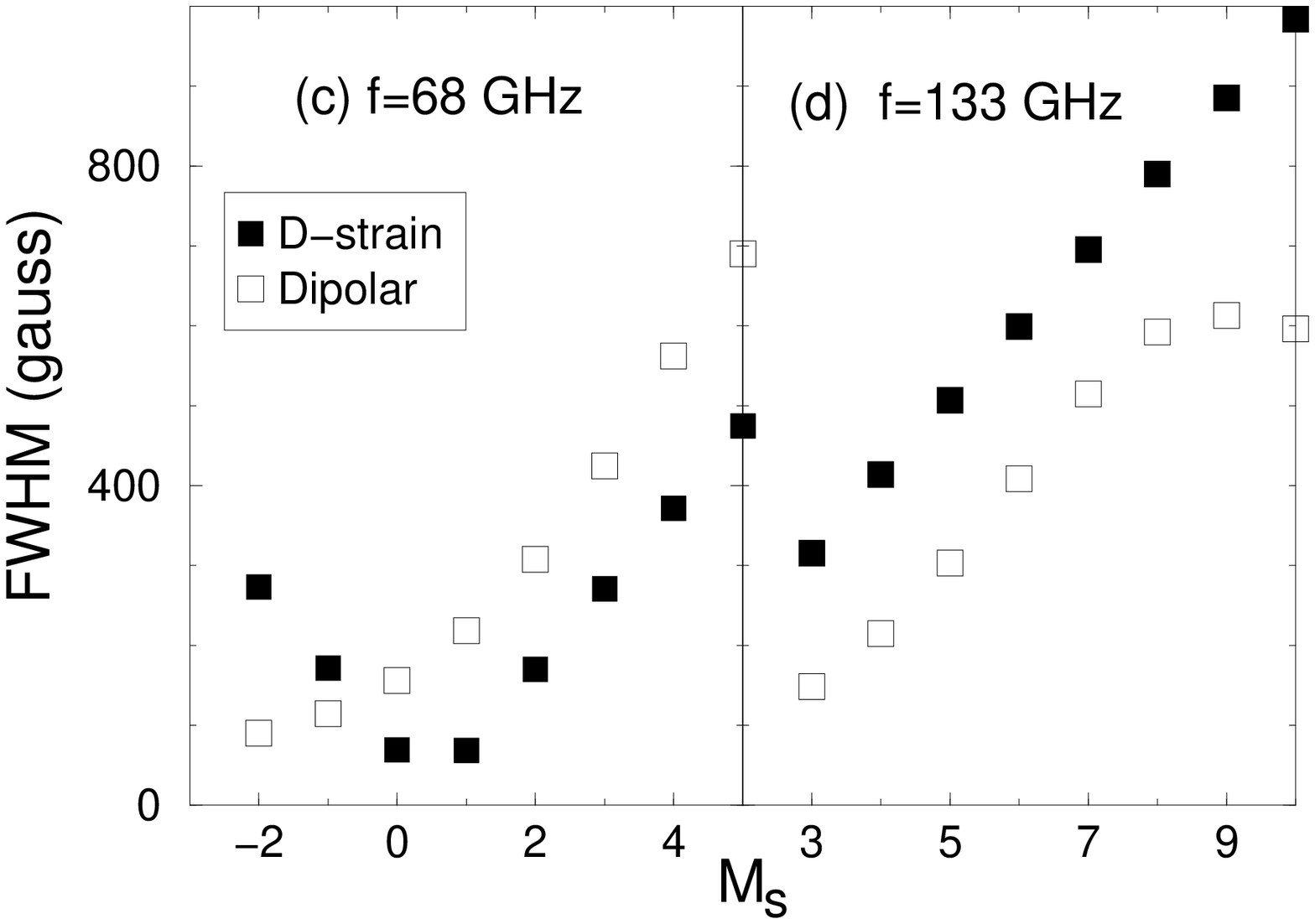}
\vspace{0.3cm}
\caption{(a) and (b)
Experimental (filled circles) and theoretical (open circles) 
Full Width at Half Maximum (FWHM) vs $M_s$ for different frequencies 
at $T$$=$$10~{\mathrm K}$ for ${\mathrm Fe_8}$.
(c) and (d) Line broadening due to the $D$-strain (filled squares)
and dipolar interactions (open squares) for (a) and (b), respectively.
Here $\vec{H} \, || \, \hat{z}$.
The fit is best when the standard deviation of $D$ is $0.01D$ and 
the effective distance between dipoles is $12~{\mathrm \AA}$.}
\label{FWHM}
\end{center}
\end{figure}

Recently, multi-frequency Electron Paramagnetic Resonance (EPR)
measurements\cite{HILL98,MACC01,PERE98} on single-crystals of the
molecular magnets Fe$_8$ and Mn$_{12}$ showed interesting
results in the linewidths as functions of the energy level $M_s$,
excitation frequency, and temperature when a magnetic field along 
the easy axis was swept with the excitation 
frequency fixed. Figures~\ref{FWHM}(a) and (b) show the 
experimental linewidths as a function of the value of $M_s$ from which the
spin system is excited. For example, $M_s$$=$5 in Fig.~\ref{FWHM}(a)
denotes the transition $M_s$$=$5$\rightarrow$4. 
The experimental results are that (i) the linewidths are about 200 G 
to 1400 G at $T$$=$10 K for ${\mathrm Fe_8}$
[Figs.~\ref{FWHM}(a) and (b)], and about 
1000 G to 2000 G at $T$$=$25 K for ${\mathrm Mn_{12}}$;
(ii) the linewidths increase {\it non-linearly} as a function of 
the absolute value of the energy eigenstate $M_s$
[Figs.~\ref{FWHM}(a) and (b)];
(iii) the linewidths attain a minimum at 
$M_s$$=$1 and $M_s$$=$0 
[Fig.~\ref{FWHM}(a)]; (iv) the linewidths decrease with increasing
frequency (compare the linewidth for $M_s$$=$5
in Fig.~\ref{FWHM}(a) with that in Fig.~\ref{FWHM}(b)). 

In this paper, we present the theoretical EPR linewidths for
single crystals of Fe$_8$ and Mn$_{12}$,
using density-matrix equations with the assumption that the uniaxial 
crystal-field anisotropy parameter $D$ and the Land{\'e} $g$ factor 
are randomly distributed around their mean values 
(``$D$-strain'' and ``$g$-strain'' effects \cite{PILB})
due to possible random defects and impurities in the samples.
We find that the calculated linewidths \cite{PARK01PRB} agree well with 
recent experimental data \cite{MACC01,PARK01PRB} as functions of the energy
level $M_s$ and the frequency.
We also briefly consider the effect of the temperature on the linewidths.
Finally we present a preliminary analysis of how
local rotations of the magnetic anisotropy axes, caused by defects 
in the samples, affect the lineshapes.

To obtain the linewidths for ${\mathrm Fe_8}$, 
we consider a single-spin system with $S$$=$$10$ in a weak oscillating
transverse field. We choose the easy axis to be the $z$-axis.  
Since the ${\mathrm Fe_8}$ clusters have an approximate $D_2$ symmetry,
the lowest-order ground-state single-spin Hamiltonian is \cite{AABB}
\begin{eqnarray}
{\cal H}_{0} &=& -D S_z^2 - E(S_x^2 - S_y^2) 
- g \mu_B H_z S_z ~, \label{eq:ham}
\end{eqnarray}
where $D$$=$$0.288k_B$ and the transverse 
anisotropy parameter, $E$$=$0.043$k_B$.\cite{MACC01}
Here $S_{\alpha}$ is the $\alpha$-th component of 
the spin operator, $g$ is 
the Land\'{e} $g$-factor ($\approx$2), $\mu_B$ is the Bohr magneton,
and $H_z$ is the longitudinal static applied field.
When $H_z$ is large enough, the eigenvalue of $S_z$, $M_s$, is a good
quantum number.
Next we introduce an interaction $V(t)$$=$$V_0 \cos(\omega t)$
between the spin system and an
oscillating transverse field $H_x$ with angular frequency 
$\omega$$\equiv$$ 2 \pi f$, where $V_0$$\propto$$H_x S_x$.
We treat $V(t)$ as a small perturbation to ${\cal H}_0$.
The interaction between the spin system and the surrounding
environment can be understood by density-matrix equations.\cite{BLUM}

We consider the case that the frequency $\omega$
is fixed while $H_z$ is quasi-statically varied to induce a resonance.  
With the selection rule $\Delta M_s$$=$$\pm 1$, 
solving the density-matrix equations for the population change with time
in the level $M_s$ due to $V(t)$, 
to first order in $V_0$ near resonance,
provides the power absorption between the levels $M_s$ and $M_s-1$:
\cite{BLUM}
\begin{eqnarray}
\!\!\frac{d{\cal E}}{dt} \! &=& \! 
\frac{({\cal E}_{M_s-1} \!\! - \!\! {\cal E}_{M_s})}{\hbar^2} {\tilde V}^2 
\Delta (\rho_{M_s,M_s} \!\! - \!\! \rho_{M_s-1,M_s-1}),  
\label{eq:dedt}\\
\!\!\Delta \!&\equiv&\! \frac{\hbar^2 \gamma_{M_s-1,M_s}}
{ (g \mu_B)^2 ( H_z - H_{\rm res})^2 + ( \hbar \gamma_{M_s-1,M_s} )^2 },
\nonumber \\ 
\!\!H_{\rm res} \!&\equiv&\!
\frac{\hbar \omega - D(2 M_s - 1)}{g \mu_B} \;.
\label{eq:res_field} 
\end{eqnarray}
Here ${\mathcal E}_{M_s}$ is the energy of the level $M_s$,
${\tilde V}$$\equiv$$|\langle M_s|V_0|M_s-1 \rangle|$,
$\rho(t)$ is the density matrix of the spin system. 
The subscripts represent eigenstates of 
the longitudinal part of ${\cal H}_0$, $\rho(t)_{m^{\prime} m}$
$=$$\langle m^{\prime}| \rho(t) | m \rangle$, 
$\gamma_{m^{\prime},m}$$=$$(W_m + W_{m^{\prime}})/2$,
$W_m$$=$$\sum_{k \neq m} W_{km}$, and
$W_{km}$ is the transition rate \cite{LEUE00}
from the $m$-th to the $k$-th eigenstate,
$H_{\rm res}$ is the resonant field, 
and $\hbar \gamma_{M_s-1,M_s}/g \mu_B$ gives a natural linewidth
which is about 5 to 50 G at 10 K and increases as $M_s$ decreases.
However, the experimentally observed linewidths are much larger 
than the natural linewidths and {\em decrease\/} as $M_s$ decreases 
[Figs.~\ref{FWHM}(a) and (b)].

To resolve these large discrepancies, we first assume that 
$D$ and $g$ are independent random variables
with Gaussian distributions centered at $0.288k_B$ and $2.00$, with
standard deviations $\sigma_D$ and $\sigma_g$, respectively.  
Then we calculate the average power absorption at a fixed frequency 
and $T$$=$$10~{\mathrm K}$ by integrating over Eq.~(\ref{eq:dedt}) 
using {\texttt{Mathematica}} \cite{MATH} to obtain 
the linewidth as a function of $M_s$.
The lineshapes depend on the magnitudes of $\sigma_D$ and $\sigma_g$,
compared to the natural linewidth determined by
$\gamma_{M_s+1,M_s}$. If $\sigma_D$ and $\sigma_g$ are much larger
than (comparable to) the natural linewidth, then the lineshape is 
Gaussian (Lorentzian).
At intermediate values of $\sigma_D$ and 
$\sigma_g$, the absorption lineshapes are neither Gaussian nor Lorentzian.
Our numerical calculations show that 
the distribution in $D$ narrows the linewidths linearly 
with decreasing absolute value of $2M_s$$-$$1$
[Figs.~\ref{FWHM}(c) and (d)], with a
slight rounding close to the linewidth minimum ($M_s$$=$1 and $M_s$$=$0)
[Fig.~\ref{FWHM}(c)].
On the other hand,
the distribution in $g$ broadens the linewidths with 
decreasing $M_s$ because the resonant field increases 
with decreasing $M_s$ [see Eq. (\ref{eq:res_field})].\cite{HILL98,PARK01PRB} 
For small-$M_s$ transitions, the lineshapes are close to a Lorentzian
because the natural linewidths are not very small compared to
the measured linewidths.

Next, we consider the effect of the dipolar interactions between
different clusters. (There is no distribution in cluster size.)
The dipolar interactions narrow the linewidths
as $M_s$ decreases because the resonant
field becomes stronger for smaller-$M_s$ transitions, and the stronger
resonant fields lead to a more polarized system 
[Figs.~\ref{FWHM}(c) and (d)]. The details of this effect on the linewidths
at a {\it fixed} temperature were reported elsewhere.\cite{PARK01PRB}
The dipolar interactions give rise to a temperature dependence of
the linewidths and a shift in the positions of the resonance lines, 
because at low temperatures ($k_B T$$\ll$$\hbar \omega$)
high energy levels are not populated. Our study shows that 
a linewidth for a particular transition increases and then 
smoothly decreases with increasing 
temperature. The maximum linewidth as a function of 
temperature moves towards lower temperatures
for larger-$M_s$ transitions, which agrees with experiments.

The competition between $D$-strain, $g$-strain, and dipolar interactions
determines the overall features of the linewidth, as a function of $M_s$.
We have varied $\sigma_D$, $\sigma_g$, and the effective distance between 
neighboring dipoles, $d$, within an experimentally acceptable range 
in order to fit the experimental data. 

For the ${\mathrm Fe_8}$ sample examined, the calculated linewidths
agree well with the experimental data at the measured frequencies
($f=68$, 89, 109, 113, 133, and 141~{GHz}) at $T$$=$$10~{\mathrm K}$,  
using $\sigma_D$$\approx$$0.01D$ and $d$$\approx$$12~{\mathrm \AA}$.
[Figs.~\ref{FWHM}(a) and (b)]
As shown in Figs.~\ref{FWHM}(c) and (d), the $D$-strain effect
and the dipolar interactions are equally important for the 
linewidths of the sample, while the $g$-strain does not 
contribute significantly (not shown). 
For Mn$_{12}$, we perform a similar analysis as above, using
the ground-state single-spin Hamiltonian\cite{AABB}
\begin{eqnarray}
{\cal H}_{0} &=& -D S_z^2 - C S_z^4 - g \mu_B H_z S_z 
\end{eqnarray}
with $D$$=$$0.55k_B$, $C$$=$$1.17 \times 10^{-3}k_B$,
and $g$$=$$1.94$.\cite{BARR97}
Detailed analysis can be found in the literature.\cite{PARK01PRB}
For this sample, the $D$-strain and $g$-strain effects play 
significant roles in the linewidths, while the dipolar interactions
are not as important as for the Fe$_8$ sample.

Next we discuss how the distribution of the directions
of the magnetic anisotropy axes of clusters affect 
the EPR lineshapes. Because of defects in the samples,
each cluster sees a slightly different crystal field 
due to the surrounding clusters,
compared to the situation in a perfect crystal. We assume that
these slightly different crystal fields result in local rotations 
of the magnetic anisotropy axes of some of the clusters
by an angle $\theta$ from the crystal $c$ axis.
The angle $\theta$ is assumed to have a Gaussian distribution 
about zero with a small standard deviation.
Hereafter $a$, $b$, and $c$ denote the crystal axes,
while $x$, $y$, and $z$ denote the magnetic anisotropy axes 
of a single cluster.

As a preliminary study, we examine the lineshapes for 
Mn$_{12}$ when the magnetic field is 
applied along the $c$ axis and the magnetic anisotropy
easy axis of a single cluster is tilted by $\theta$ from
the $c$ axis.
Then the single-spin Hamiltonian, in terms of the spin 
operators along the magnetic anisotrpy axes, becomes
\begin{eqnarray}
{\cal H}&=&-D S_z^2 - g \mu_B H S_z \cos \theta 
+ g \mu_B H S_x \sin \theta \:.
\label{eq:ham_hz_xyz} 
\end{eqnarray}
For simplicity, we drop the fourth-order anisotropy terms.
We also set $\psi=90^{\circ}$, where $\psi$ is the third of the 
Euler angles defined in literature,\cite{GOLD} which does not affect
the eigenvalues of ${\cal H}$.

\begin{figure}
\begin{center}
\epsfxsize=3.4cm
\epsfysize=3.cm
\epsfbox{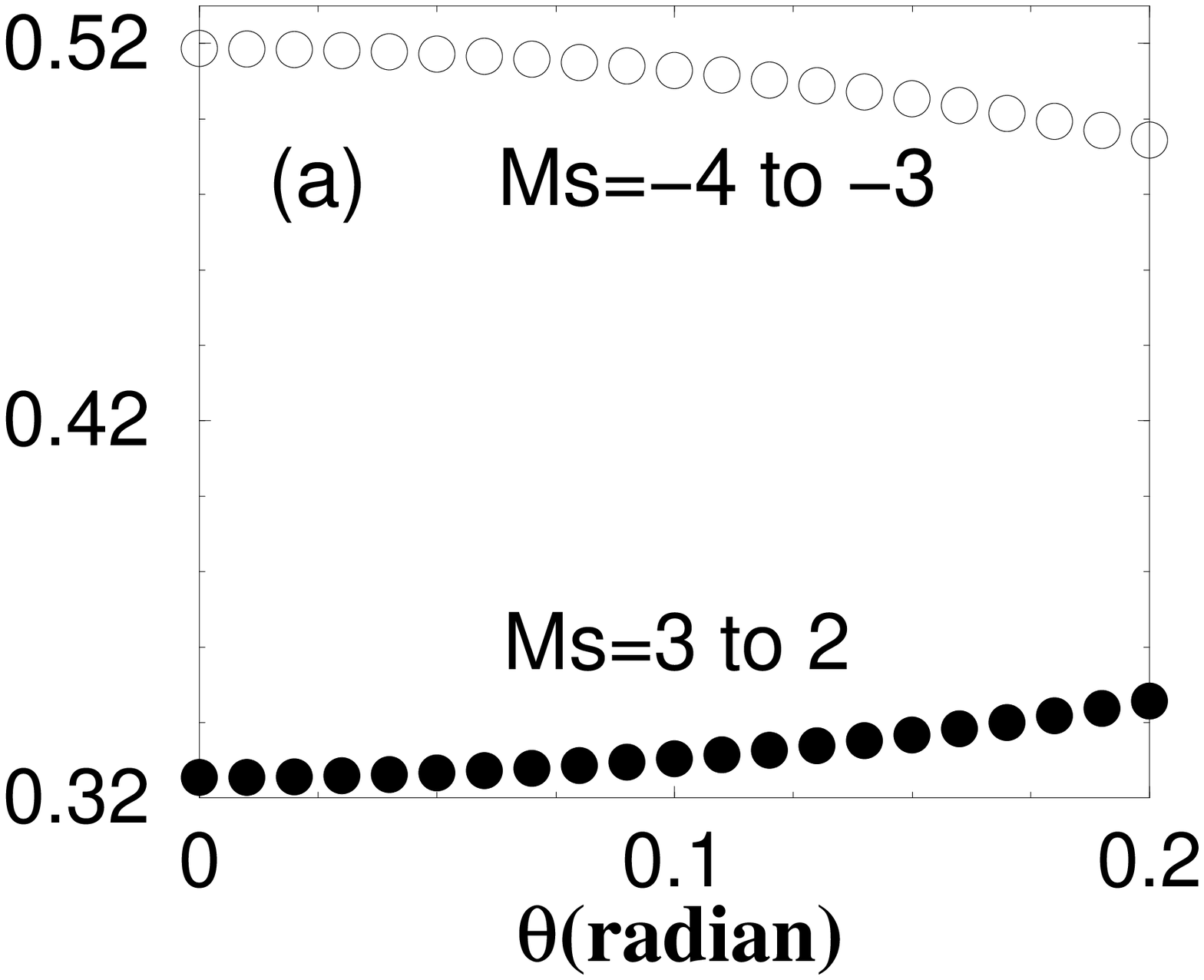}
\hspace{0.5cm}
\epsfxsize=3.4cm
\epsfysize=3.cm
\epsfbox{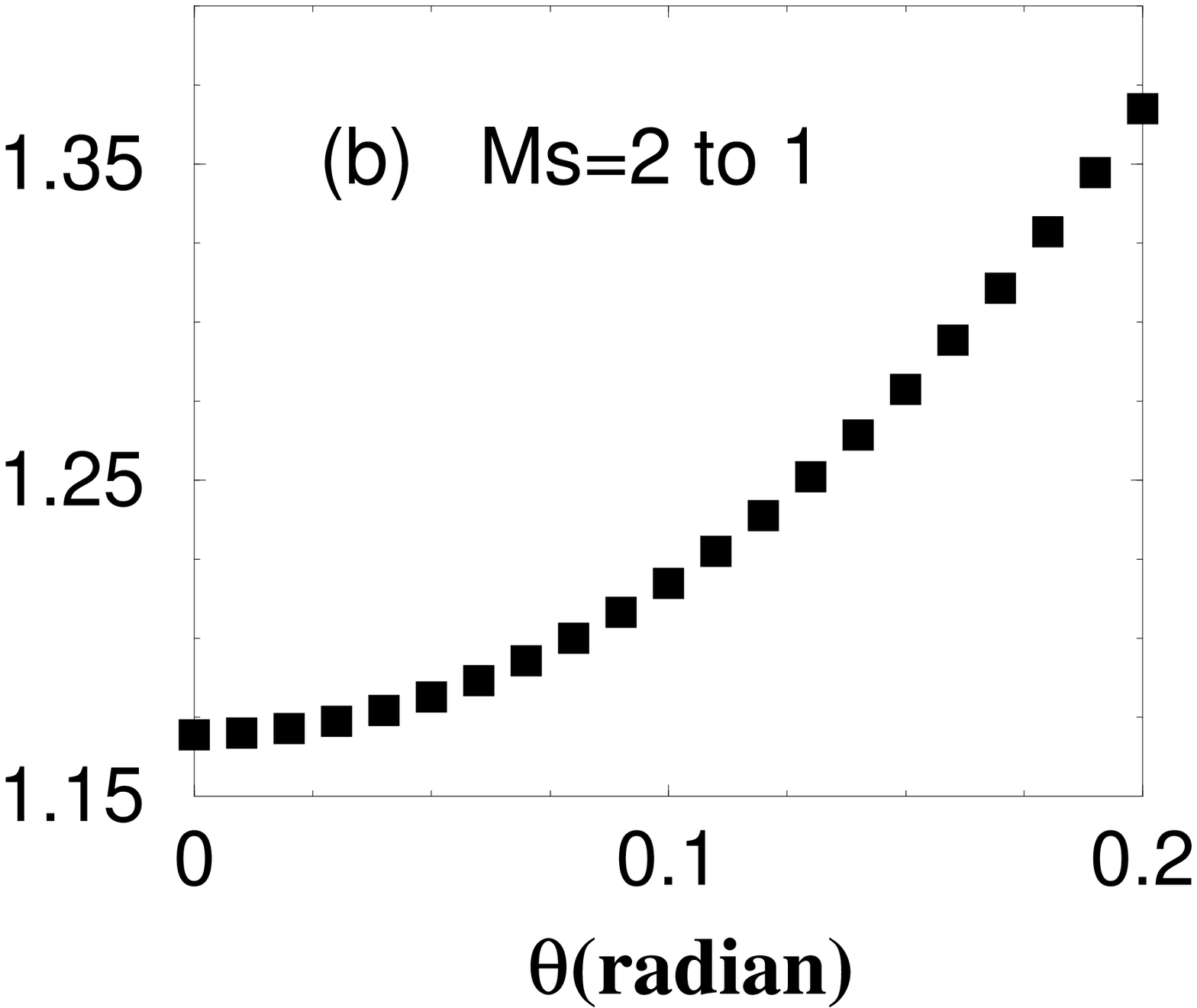}
\vspace{0.3cm}
\caption{The resonant field vs $\theta$
for the transitions (a) $M_s$$=$$-4$$\rightarrow$$-3$ 
and $M_s$$=$3$\rightarrow$2, and (b) $M_s$$=$2$\rightarrow$1 
at $f$$=$66.135 GHz for $\vec{H} \, || \, \hat{c}$.
Each curve is symmetric about $\theta$$=$0.}
\label{shape}
\end{center}
\end{figure}

Because of the local rotations of the magnetic anisotropy axes,
some of the clusters experience slightly different resonant fields
than those aligned with the crystal $c$ axis.
In contrast to the Gaussian distributions of the resonant field
due to the distributions in $D$ and $g$, this effect gives rise
to an asymmetry in the resonant field, which leads to an asymmetry
in the EPR lineshapes. Assuming that $\theta$ is small, we 
treat the terms proportional to $\sin \theta$ as small perturbations 
to the rest of the terms in Eq.~(\ref{eq:ham_hz_xyz}). 
Using second-order perturbation theory, we obtain the resonant fields
as a function of $\theta$, shown in Fig.~\ref{shape}. 
Each curve in Fig.~\ref{shape} is symmetric about $\theta$$=$0.
Figure~\ref{shape} shows that for transitions between 
$M_s$ and $M_s$$+$1 ($M_s$$=$$-4$$\rightarrow$$-3$) an asymmetry 
appears in the direction of {\it decreasing} field, while for transitions 
between $M_s$ and $M_s$$-$1 ($M_s$$=$3$\rightarrow$2, $M_s$$=$2$\rightarrow$1)
the asymmetry is in the direction of {\it increasing} field.
At the examined frequency ($f$$=$66.135 GHz)
the asymmetry effect is more significant for the small-$M_s$ transitions.
Based on the analytic form of the resonant field, we can obtain 
analytically the distribution of the resonant field
at a particular frequency for transitions between the levels 
$M_s$ and $M_s$$\pm$1, including Gaussian distributions 
of $D$, $g$, and $\theta$.\cite{PARKPRP} For simplicity, 
we have neglected the effects of natural linewidths, 
dipolar interactions, and temperature on the EPR lineshapes.

In conclusion, we have examined the EPR linewidths as functions of
$M_s$, frequency, and temperature for single crystals of the molecular 
magnets Fe$_8$ and Mn$_{12}$. We found that the distribution in $D$
is important to explain the linewidths for both of the molecular magnets,
and that the dipolar interactions are crucial to understand the temperature 
dependence of the linewidths. Our preliminary study of the EPR lineshapes
shows that the distribution of the directions of the magnetic 
anisotropy axes of clusters provides an asymmetry in the EPR spectra.



Funded by NSF Grants DMR-9871455, DMR-0120310, and 
DMR-0103290, and FSU-CSIT, FSU-MARTECH, and Research Corporation.

\end{document}